\documentclass[twocolumn]{article}          
\usepackage[super,sort&compress,comma]{natbib}
\usepackage[tight]{subfigure}
\usepackage{framed,multirow}
\usepackage{makecell}
\usepackage{amsmath}
\usepackage{fancyhdr}		
\usepackage{xcolor}

\newcommand{\captionv}[3]{\begin{center}\parbox{#1cm}{\caption[#2]{{\sf #3}}}
        \end{center}}

\usepackage[section]{placeins}   %

\usepackage{graphicx}

\makeatletter \renewcommand\@biblabel[1]{$^{#1}$} \makeatother
\newcommand{\cen}[1]{\begin{center} #1 \end{center}}

\usepackage[mathlines,switch]{lineno}

\usepackage{hyperref}
\hypersetup{ colorlinks,
    citecolor=blue,
    filecolor=blue,
    linkcolor=blue,
    urlcolor=blue
}

\usepackage{xcolor}

\definecolor{gray}{rgb}{0.6,0.6,0.6}
\definecolor{red}{rgb}{0.85,0,0}
\definecolor{green}{rgb}{0,0.85,0}
\definecolor{blue}{rgb}{0,0,0.85}
\definecolor{beige}{rgb}{0.92,0.87,0.78}
\usepackage[all]{hypcap}    

\begin{document}

\onecolumn

\cen{\sf {\Large {\bfseries Fiducial marker recovery and detection from severely truncated data in navigation assisted spine surgery} \\
\vspace*{10mm}
Fuxin Fan} \\
Pattern Recognition Lab, Friedrich-Alexander-Universit\"at Erlangen-N\"urnberg, Erlangen 91058, Germany
\vspace{5mm}\\
{\Large Bj\"orn Kreher}\\
Siemens Healthcare GmbH, Forchheim, Germany
\vspace{5mm}\\
{\Large Holger Keil}\\
Department of Trauma and Orthopedic Surgery, Universit\"atsklinikum Erlangen, Friedrich-Alexander-Universit\"at Erlangen-N\"urnberg, Erlangen 91054, Germany
\vspace{5mm}\\
{\Large Andreas Maier}\\
Pattern Recognition Lab, Friedrich-Alexander-Universit\"at Erlangen-N\"urnberg, Erlangen 91058, Germany
\vspace{5mm}\\
{\Large Yixing Huang}\\
Department of Radiation Oncology, Universit\"atsklinikum Erlangen, Friedrich-Alexander-Universit\"at Erlangen-N\"urnberg, Erlangen 91054, Germany
\vspace{5mm}\\
Version typeset \today\\
}

\pagenumbering{roman}
\setcounter{page}{1}
\pagestyle{plain}
Correspondence: Yixing Huang, Department of Radiation Oncology, Universit\"atsklinikum Erlangen, Friedrich-Alexander-Universit\"at Erlangen-N\"urnberg, Erlangen 91054, Germany. Email: yixing.yh.huang@fau.de\\

\begin{abstract}
\noindent {\bf Purpose:} Fiducial markers are commonly used in navigation assisted minimally invasive spine surgery and they help transfer image coordinates into real world coordinates. In practice, these markers might be located outside the field-of-view (FOV) of C-arm cone-beam computed tomography (CBCT) systems used in intraoperative surgeries, due to the limited detector sizes. As a consequence, reconstructed markers in CBCT volumes suffer from artifacts and have distorted shapes, which sets an obstacle for navigation. \\
{\bf Methods:} In this work, we propose two fiducial marker detection methods: direct detection from distorted markers (direct method) and detection after marker recovery (recovery method). For direct detection from distorted markers in reconstructed volumes, an efficient automatic marker detection method using {two neural networks and a conventional circle detection algorithm} is proposed. For marker recovery, {a task-specific learning strategy is proposed to recover markers from severely truncated data}. {Afterwards,} a conventional marker detection algorithm {is applied} for position detection.\\
{\bf Results:} The two methods are evaluated on simulated and real data. The direct method achieves 100\% detection rates with maximal 2-pixel difference on simulated data with normal truncation and simulated data with severe noise, but fails to detect all the markers in extremely severe truncation case. The recovery method detects all the markers successfully with maximal 1 pixel difference on all simulated data sets. For real data, both methods achieve 100\% marker detection rates with mean registration error below 0.2 mm.\\
{\bf Conclusions:} Our experiments demonstrate that the direct method is capable of detecting distorted markers accurately and the recovery method with task-specific learning has high robustness and generalizability on various data sets. 
The task-specific learning is able to reconstruct structures of interest outside the FOV from severely truncated data, which has the potential to empower CBCT systems with new applications.
\end{abstract}

\newpage     

\tableofcontents

\clearpage
\twocolumn
\pagenumbering{arabic}
\setcounter{page}{1}
\pagestyle{fancy}
\section{Introduction}
Spine is one of the most important parts of the human body. It supports our trunk and allows us to move {upright} and bend freely. However, due to accidents or chronic degeneration, many people suffer from spine disorders and need spine surgeries. Minimally invasive spine surgery (MISS) is an important surgical technique resulting in less collateral tissue damage, bringing measurable decrease in morbidity and more rapid functional recovery than conventional open surgery techniques \cite{pmid21160389}. Among the numerous forms of MISS, {navigation techniques play} an essential role \cite{Vaishnav2019}. With accurate registration between the image and the real world coordinate systems \cite{article}, {a} navigation system {visualizes} surgical equipment on a monitor, assisting surgeons to perform precise operations on patients. It {allows} more accurate pedicle screw placement compared to conventional surgical techniques \cite{Tian2009,Virk2019}. And it reduces the amount of X-ray exposure to surgeons and patients as well \cite{pmid21304425}.

\begin{figure*}[hbt!]
    \centering
    \begin{minipage}[t]{0.24\linewidth}
    \subfigure[C-arm system]{
    \includegraphics[width=\linewidth]{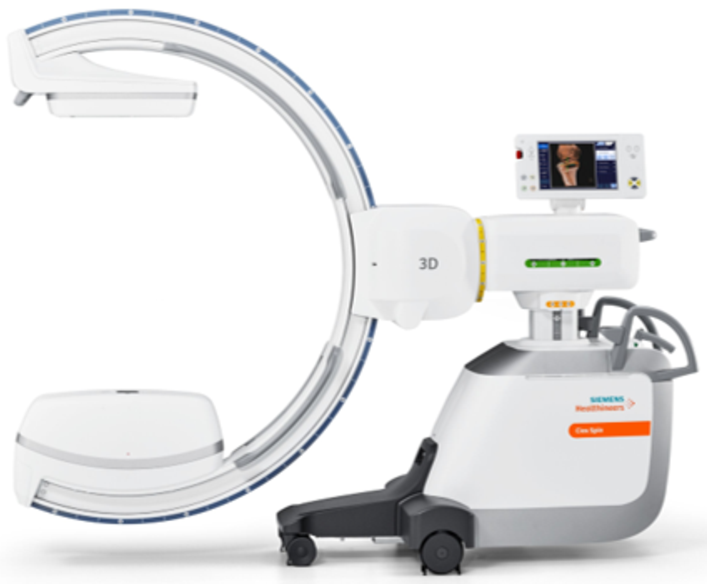}
    \label{subfig:Intro_Carm}
    } 
    \end{minipage}    
    \begin{minipage}[t]{0.2\linewidth}
    \subfigure[Projection after log transform]{
    \includegraphics[width=\linewidth]{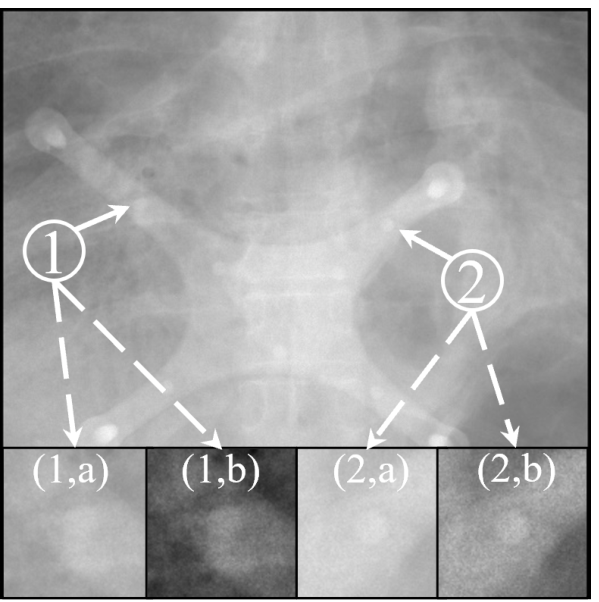}
    \label{subfig:Intro_Project}
    }
    \end{minipage}
    \begin{minipage}[t]{0.2\linewidth}
    \subfigure[Markers near FOV]{
    \includegraphics[width=\linewidth]{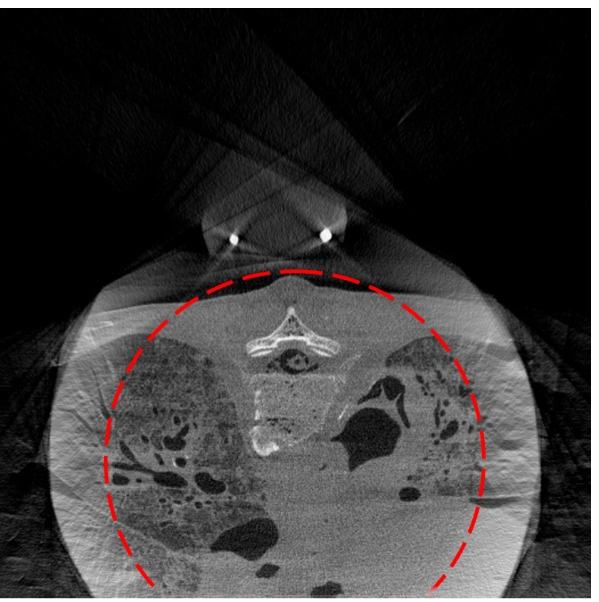}
    \label{subfig:Intro_ReconNear}
    }
    \end{minipage}
   \begin{minipage}[t]{0.2\linewidth}
    \subfigure[Markers far from FOV]{
    \includegraphics[width=\linewidth]{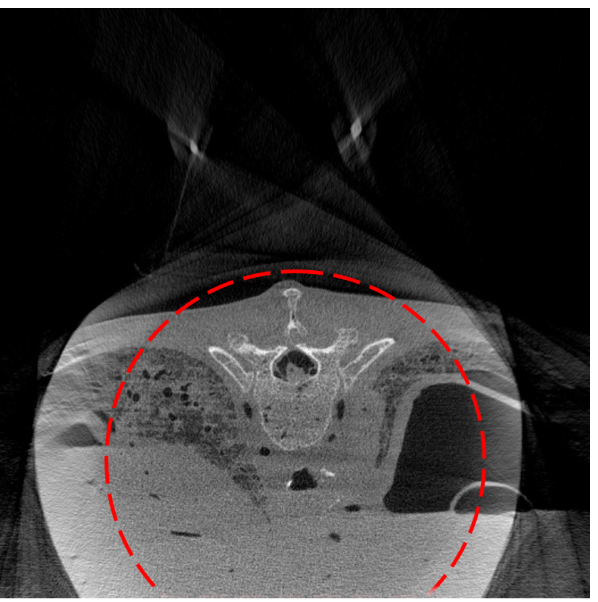}
    \label{subfig:Intro_ReconFar}
    }
    \end{minipage} 
   \captionv{16.5}{}{A C-arm CBCT system and its image examples for fiducial markers in projection domain and image domain: (a) A C-arm system from Siemens Healthcare$^{\copyright}$; (b) One projection with markers, {where} (1,\,a) and (1,\,b) are the same marker but {hardly visible even though displayed} with different {intensity windows, while the other marker is slightly better visualized in (2,\,a) and (2,\,b)}; (c) One CT {slice} with fiducial markers close to {the} patient's back, window: {[-1000, 500]\,HU}; (d) One CT slice with fiducial markers far away from {the} patient's back, {where severer marker distortion is observed,} window: [-1000, 500]\,HU. The FOV is indicated by the dashed circle in (c) and (d).
   \label{DeformedMarkers}}
\end{figure*}

C-arm {cone-beam computed tomography (CBCT)} systems are widely used in navigation assisted spine surgery. Due to their limited detector size, reconstructed CBCT volumes have a field-of-view (FOV), {which is not large enough to cover spine and fiducial markers together. In intraoperative spine surgeries, transferring patients into a multi-slice CT system with a large FOV not only delays treatment procedures but also introduces difficulty in marker registration because of anatomy changes between two imaging systems. Assembling large detectors or moving the isocenter close to the detector leads to bulky C-arms or small space for operations. In both cases, the dose exposure to patients is increased. Therefore, in practice the  target spine region is placed inside the FOV, while markers are outside the FOV} which are distorted due to missing data. {Especially for obese patients, marker distortion is a common problem since the X-ray markers are far away from the spine.} In Fig.~\ref{DeformedMarkers}, a C-arm system and {some marker image examples} are displayed. {Fig.\,\ref{subfig:Intro_Carm} is a mobile C-arm CBCT system.} Fig.~\ref{subfig:Intro_Project} is one projection image {after log transform} containing seven markers and two of them are {zoomed-in for better visualization}. Due to the obstacle of ribs, marker {No.\,1} in (1,a) and (1,b) is not {distinguishable} even with {contrast adjustment}. 
Thus it is difficult to detect markers in the projection domain. Fig.~\ref{subfig:Intro_ReconNear} and ~\ref{subfig:Intro_ReconFar} are {two exemplary} slices from reconstructed CBCT volumes and the red circles {indicate the FOV boundaries}. The {marker holders} in Fig.~\ref{subfig:Intro_ReconNear} and ~\ref{subfig:Intro_ReconFar} {have} 1 cm and 5 cm distances away from the {patients' backs, respectively}. It is clear that the further the markers are away from the spine, the severer deformation in these markers will occur, since more projection data is missing for the markers. Besides, the intensity of the markers is decreasing with the increasing distance.

Under {the circumstance of distorted markers}, the navigation system with conventional {marker} detection {algorithms typically fails} to detect the correct positions of these markers, thus causing failure in the {subsequent} registration process. Therefore, one purpose of this work is to develop an algorithm to detect positions directly from the distorted markers.
\begin{figure*}[hbt]
   \centering
   \includegraphics[width=0.7\linewidth]{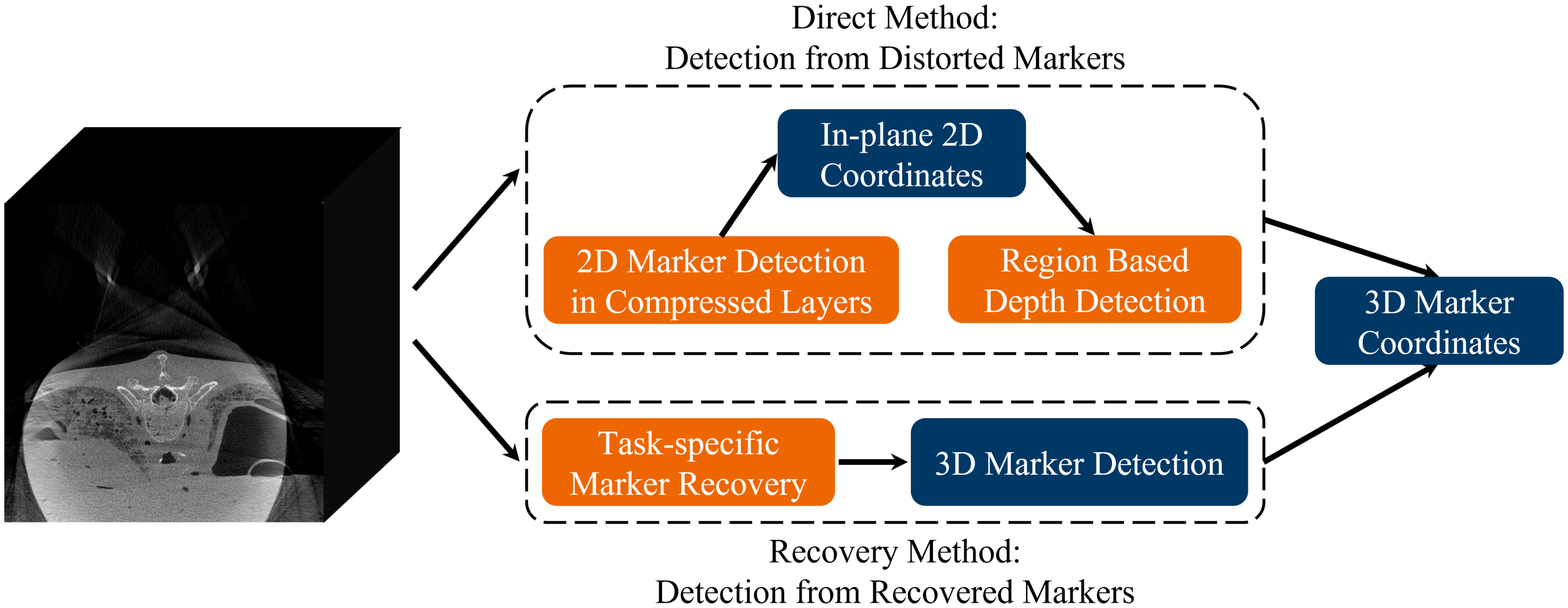}
   \captionv{16.5}{}
   {
   A brief sketch of the two algorithms for marker detection from severely truncated data, where the orange blocks are our main contributions. 
   \label{GeneraliPipeline}
    }  
\end{figure*}

{Alternatively, distorted markers need to be restored so that existing marker detection algorithms from different vendors can be simply reused for recovered markers. Therefore, the other purpose of this work is} to develop a universal marker recovery algorithm to recover both the shape and the intensity of markers from different vendors in reconstructed volumes. With such a recovery algorithm, one single CBCT system is capable to provide image guidance for different navigation systems from different vendors. {The marker recovery problem is fundamentally a truncation correction problem. However, due to the severe truncation level, state-of-the-art deep learning algorithms \cite{fonseca2021evaluation,huang2021data} cannot achieve satisfying performance. In addition, the instability of deep learning algorithms in heterogeneous clinical data is a major concern in practical applications \cite{huang2018some,antun2020instabilities}. To gain robustness and generalizability, a task-specific learning strategy is proposed to help neural networks focus on structures of interest (SOI) only.
  }
 
A brief sketch of the two above mentioned approaches is displayed in Fig.~\ref{GeneraliPipeline}. The main contributions of this work are summarized as follows:
 
1. {Multi-step} coordinate determination {directly from distorted markers}: 
{Two neural networks, as well as a conventional 2D marker detection algorithm, are used to detect in-plane and depth position information respectively. To the best of our knowledge, this algorithm is the first deep learning-based algorithm for detecting distorted markers in CBCT volumes.}
{With the multi-step detection mechanism, it is robust to eliminate false positive (FP) cases.}

2. Marker recovery: {our work is} the first application of deep learning in marker recovery for universal automatic navigation systems and the network has generalizability in real CBCT volumes with different patterns: thoracic and lumbar volumes, markers from different vendors, and the existence of K-wires.

3. Task-specific learning: a special data preparation strategy {for SOI reconstruction from severely truncated data}, which helps the {neural network} focus on the SOI only and maintain robust.

\section{Related work}
\subsection{Marker detection}
{Fiducial markers are widely used in various medical applications and many marker detection algorithms have been developed. Among them, template matching is commonly used in practice \cite{fledelius2014real,bertholet2017fully,campbell2017automated,mylonas2019deep}. With prior information on markers like geometric shapes, intensity, and spatial distribution, a template for fiducial markers is constructed to search for matching patterns in images. Such template matching algorithms typically work reliably for undistorted markers. However, deformation upon implantation or image artifacts caused by missing data degrades their performance considerably \cite{mylonas2019deep}.}

{As spherical markers are the most widely used fiducial markers, Hough transform \cite{duda1972use} plays a very important role in marker detection}. It has many variants such as standard \cite{ballard1981generalizing}, probabilistic \cite{kiryati1991probabilistic}, fast \cite{ogundana2007fast} and combined multi-point \cite{camurri20143d} Hough transform. It succeeds in sphere detection in medical imaging, including glenohumeral joint detection in MRI and CT \cite{van2002determination}, the detection of glomeruli in MRI \cite{xie2012magnetic}, and hip joint analysis in CT \cite{lee2019three}. {Like template matching, Hough transform-based methods work reliably only on undistorted markers.}
 
{With the success of deep learning in numerous applications, it has also been applied to fiducial marker detection.}
Mylonas et al. \cite{mylonas2019deep} proposed a 2D {convolutional neural network (CNN)} based real-time multiple object tracking system with sliding windows to classify and segment fiducial markers. Gustafsson et al. \cite{gustafsson2020development} proposed a HighRes3DNet model to segment spherical gold fiducial markers in MRI. Nguyen et al. \cite{nguyen2020beadnet} detected spherical markers by {the} BeadNet, which shows smaller detection error and lower variance {than} conventional Hough transform based methods. To detect mutli-category fiducial markers in CT volumes, Regodic et al. \cite{regodic2021automated} proposed a three-step hybrid approach with a 3D CNN, achieving correct classification rates of 95.9\% for screws and 99.3\% for spherical fiducial markers respectively. {Although the above mentioned deep learning approaches are investigated for markers without distortion, they have the risk of FP detection. For example, the hybrid approach \cite{regodic2021automated} has FP rates of 8.7\% and 3.4\% for screws and spherical fiducial markers, respectively. In this work, the fiducial markers reconstructed from severely truncated data suffer from shape deformation and intensity decrease, making} it difficult for the above approaches to detect them accurately.

\subsection{Truncation correction}

{Data truncation is a common problem for CBCT systems with flat-panel detectors because of equipped collimators for dose reduction or their limited detector sizes. In navigation assisted spine surgery, the deformation and the insufficient intensity of markers are also caused by data truncation. Therefore, truncation correction is beneficial to restore markers.}
A major category of approaches for truncation correction are based on heuristic extrapolation. They tend to extend the truncated {projection data, e.g.,} by symmetric mirroring \cite{2000MedPh..27...39O},  cosine function fitting \cite{pmid15702338} and water cylinder extrapolation (WCE) \cite{https://doi.org/10.1118/1.1776673}. Other categories work on the image reconstruction process, like the differentiate back-projection \cite{NooSep2004} and decomposing the ramp filter into a local Laplace filter and a nonlocal low-pass filter \cite{6670089}. 
{In addition, compressed sensing techniques were widely applied \cite{yu2009compressed,yang2010high}.}

{Recently, deep learning methods have been applied to FOV extension, achieving impressive results. Fourni\'e et al. \cite{fournie2019ct} have applied the U-Net \cite{ronneberger2015unet} to postprocess images reconstructed from linearly extrapolated data. Fonseca et al. \cite{fonseca2021evaluation} proposed a deep learning based algorithm called HDeepFov, which has a better performance than the latest commercially released algorithm HDFov. Huang et al. \cite{10.1007/978-3-658-29267-6_40} provided a data consistent reconstruction method for FOV extension combining the U-Net and iterative reconstruction with total variation \cite{huang2018scale}. This method is extended to a general plug-and-play framework \cite{huang2021data}, where various deep learning methods and conventional reconstruction algorithms can be plugged in. 
It guarantees the robustness and interpretability for structures inside the FOV. However, the structures outside the FOV relies strongly on the performance of the neural network. In the application of marker recovery for navigation assisted spine surgery, the markers suffer from severe truncation and are located outside the FOV. Therefore, all the above mentioned deep learning methods have limited performance for this application.
}

\section{Materials and Methods}
In this section, we introduce the contents of Fig.~\ref{GeneraliPipeline} in detail.
\subsection{Direct method: detection from distorted markers}
\begin{figure*}[t]
   \centering
   \includegraphics[width=0.8\textwidth]{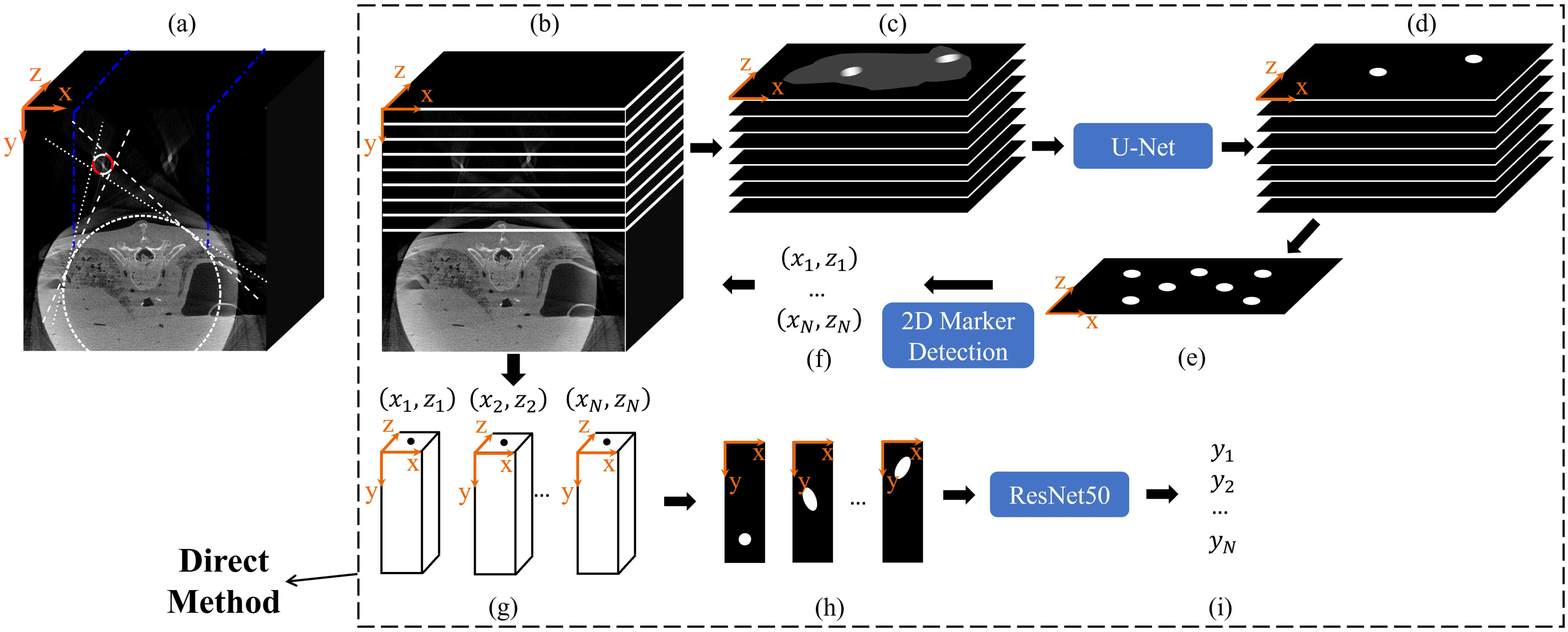}
   \captionv{16.5}{}
   {The direct method (in box) with {two 2D neural networks and a conventional 2D marker detection algorithm}, where the U-Net {segments} the markers in {the x-z plane, the conventional 2D marker detection algorithm extracts the in-plane coordinates,} and {the} ResNet50 locates the depth information for markers. {In (a), the white solid circle illustrates an ideal marker boundary in the x-y plane, which is larger {than a normal marker} for better visualization. There are four tangent lines connecting the marker boundary and the FOV boundary. The two points of tangency from the two white dotted lines determine a fragment (marked by red) on the left side of the marker boundary, while those from the two white dashed lines determine the other fragment (marked by red) on the right side. The two red fragments can be reconstructed, while the remaining white fragments are distorted.} The markers in practice are within the region between the blue dashed lines. {(b)-(i) are different steps of the direct method:} (b) The volume with flat slabs; (c) The integrated layers {from previous flat slabs}; (d) Predictions by the U-Net; (e) The layer compressed from previous predictions; (f) 2D in-plane coordinates; (g) Long cuboids from volume; (h) Integrated rectangular images from long cuboids; (i) Depth information given by the ResNet50.
   \label{Network2}
    }  
\end{figure*}
The direct method tries to locate the accurate positions of distorted markers directly in CBCT volumes reconstructed from severely truncated data. Considering the {expensive} computation of 3D {networks}, a {multi-step coordinate} determination method combining two 2D networks {and a conventional 2D marker detection algorithm} is proposed.

{As displayed in Fig.\,\ref{Network2}, the image coordinate system is defined as follows:} the x-y plane is the transverse plane, the y-z plane is the sagittal plane, and the x-z plane is the frontal plane, considering the correspondence to anatomical planes for human body. 
With the prior information that markers are {fixed inside a fiducial holder} and it is placed above a patient's back (typically between the two blue lines in Fig.\,\ref{Network2}(a)), the markers are not overlapping each other along {the frontal axis, i.e., the y-axis}. {The distortion of markers outside the FOV fundamentally originates from limited angular X-rays passing through them. In limited angle tomography \cite{quinto2007local,huang2016image}, boundaries tangent to available X-rays are well preserved in reconstructed images (see Fig.\,3.6 in \cite{frikel2013reconstructions}). To illustrate this, a white solid circle is drawn in Fig.\,\ref{Network2}(a) to represent an ideal marker boundary, which is larger than a normal marker for better visualization. There are four tangent lines connecting the marker boundary and the FOV boundary. The two points of tangency from the two white dotted lines determine a fragment (marked by red) on the left side of the marker boundary, while those from the two white dashed lines determine the other fragment (marked by red) on the right side. The two red fragments can be reconstructed, while the remaining white fregments are distorted. The distortion occurs along the isocenter to the marker center, which is approximately along the y direction since the markers are in the region between the two blue dashed lines. Therefore, in the x-z plane which is perpendicular to the y-axis, marker shapes are mostly preserved (please refer to the middle column of Fig.\,\ref{CombinedImages}; Fig. 5 in \cite{huang2020limited} is another example).}

The markers in the projection {image} in {Fig.\,\ref{DeformedMarkers}(b)} are not distinctive due to the integral of all the elements together. {To overcome this problem,} a higher contrast between the markers and their surroundings can be {achieved} by the integral {along a shorter ray path}.
Therefore, the upper part of the volume is equally split into 8 flat slabs in parallel to the x-z plane {(Fig.~\ref{Network2}(b))} and each of them is integrated along the y-axis into one layer (Fig.~\ref{Network2}(c)). {Note that the x-z plane is chosen for the compressive layer because the markers have little shape deformation in this plane as aforementioned. The flat slabs are integrated along the y-axis (orthogonal projection) instead of along radial directions (perspective projection) to keep marker size.}{ Afterwards, a neural network, particularly the U-Net in this work, is applied to segment maker areas in each layer (Fig.~\ref{Network2}(d)).} All the layers are further compressed into a single layer (Fig.~\ref{Network2}(e)) containing distinct 2D markers. Their 2D in-plane coordinates (Fig.~\ref{Network2}(f)) can be easily obtained by conventional 2D marker detection methods. 
It is worth noting that the in-plane coordinates are obtained by a conventional 2D marker detection algorithm instead of directly by the neural network, since prior information on the markers like radius and intensity can be integrated to remove intermediate false positives (IFPs). For the {$i$-th} marker, its 2D in-plane coordinates are denoted by $(x_i,z_i)$. Afterwards, $N$ long cuboids (Fig.~\ref{Network2}(g)) with the previous detected 2D in-plane coordinates as their central axes are selected out from the reconstructed volume. Each cuboid is integrated along {the} longitudinal axis, {i.e. the z-axis,} into a rectangular image (Fig.~\ref{Network2}(h)) and {a second neural network, particularly the ResNet50 in this work,} is trained to locate the depth information (Fig.~\ref{Network2}(i)) for {each} marker. Combining the depth information with {the} previous in-plane coordinates, the 3D positions of all the markers are obtained.

\subsection{Recovery method: detection from recovered markers}
\subsubsection{Neural Networks}
For marker recovery, we investigate two U-Net-based neural networks, FBPConvNet \cite{jin2017deep}  and Pix2pixGAN \cite{isola2017image}, which are the state-of-the-art neural networks for truncation correction \cite{huang2021data}.

\begin{figure}[hbt!]
  \centering
  \includegraphics[width=0.48\textwidth]{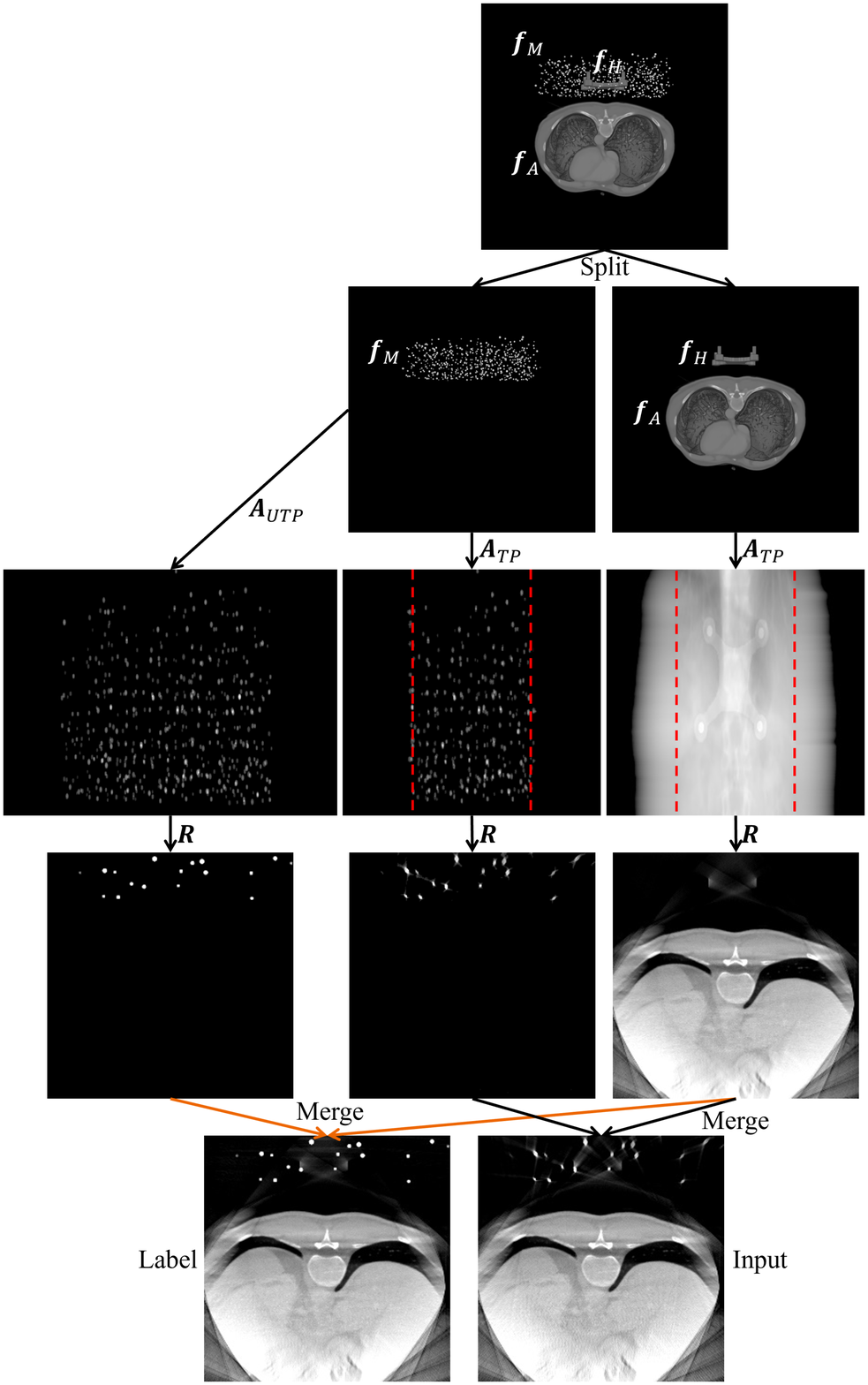}
  \captionv{8}{}
  {Task-specific data preparation for marker recovery. {The markers ($\boldsymbol{f}_{\text{M}}$) are forward projected to large ($\boldsymbol{A}_{\text{UTP}}$) and small ($\boldsymbol{A}_{\text{TP}}$) detectors to get untruncated and truncated data respectively, while the other structures (the holder $\boldsymbol{f}_{\text{H}}$ and antomical structures $\boldsymbol{f}_{\text{A}}$) are forward projected to the small detector only. The input is the combination of the markers and the others, both reconstructed ($\mathcal{R}$) from truncated data. The label of the network is the combination of the markers reconstructed ($\mathcal{R}$) from untruncated data and the others reconstructed from truncated data. Therefore, the difference between the label and the input, which the network learns, lies only in the marker areas. The vertical dashed lines mark the truncation boundaries and the projection data outside the dashed lines are extrapolated by WCE.}
  \label{Fig:dataset2}
    } 
\end{figure}


\subsubsection{Conventional data preparation}
To prepare training data, it is straightforward to use images reconstructed from truncated data as the network input and images reconstructed from untruncated (complete) data as the network output.


 For such conventional training data preparation, reference CT volumes are forward projected into a small detector for truncated projection data and a large (virtually extended) detector for complete data, respectively. 
 Afterwards, volumes for input data and label data are reconstructed respectively from truncated and complete data. 
 Such conventional data preparation for our particular marker recovery application can be represented as follows,
\begin{equation}
\begin{array}{c}
    Input = \mathcal{R} \left(\boldsymbol{A}_{\text{TP}}(\boldsymbol{f}_{\text{(A+H)}} + \boldsymbol{f}_{\text{M}})\right), \\
     Label = \mathcal{R} \left(\boldsymbol{A}_{\text{UTP}}(\boldsymbol{f}_{\text{(A+H)}} + \boldsymbol{f}_{\text{M}})\right),
\end{array}
\label{eqn:conventionalMakerPre}
\end{equation}
where $\boldsymbol{f}_{\text{(A+H)}}$ denotes anatomical structures and marker holders, and $\boldsymbol{f}_{\text{M}}$ denotes markers in Fig.\,\ref{Fig:dataset2}. The operation $\mathcal{R}$ means image reconstruction, which is FDK reconstruction from water cylinder extrapolated data in particular in this work \cite{huang2021data}; $\boldsymbol{A}_{\text{TP}}$ and $\boldsymbol{A}_{\text{UTP}}$ are truncated forward projection and untruncated forward projection operators, respectively. The sign ``$+$'' means combination here instead of the mathematical addition.

Since no truncation occurs in the large detector, all structures are well kept in the label data. With such conventional data preparation, the network is expected to recover all the imperfect regions of the input data, including markers, holders, and all the anatomical structures.

\subsubsection{Task-specific data preparation}
In the scenarios with severely truncated data, it is very challenging to restore complete structures outside the FOV accurately due to the large amount of missing data. In this work, we propose a task-specific learning strategy to let neural networks focus on reconstructed SOI only, since only certain structures outside the FOV are of interest in many applications.

In order to let neural networks focus on SOI, one potential approach is to apply more weights on such structures than others. However, in images reconstructed from truncated data, streak artifacts spread along missing angular ranges to a large extent. Therefore, it is infeasible to segment such regions for applying weights. If the weights are only applied in the SOI area, artifacts from SOI will remain outside the weighted area. To avoid such difficulty, we propose our special data preparation method for task-specific learning.

 The task-specific data preparation constrains the difference between the input and label image locate in the SOI only. The pipeline for marker recovery is shown in Fig.~\ref{Fig:dataset2}.
Instead of generating the input and label directly, the markers are projected into a small detector and a large one to obtain deformed and undistorted markers respectively. Afterwards they are merged with the volume containing anatomical structures and holders reconstructed from truncated data. Correspondingly, Eqn.\,(\ref{eqn:conventionalMakerPre}) becomes the following,
 \begin{equation}
 \begin{array}{c}
      Input = \mathcal{R} (\boldsymbol{A}_{\text{TP}}(\boldsymbol{f}_{\text{(A+H)}}) + \mathcal{R} (\boldsymbol{A}_{\text{TP}}(\boldsymbol{f}_{\text{M}})),  \\
      Label = \mathcal{R} (\boldsymbol{A}_{\text{TP}}(\boldsymbol{f}_{\text{(A+H)}}) + \mathcal{R} (\boldsymbol{A}_{\text{UTP}}(\boldsymbol{f}_{\text{M}})).
 \end{array}
\label{eqn:markerSpecificPreparation}
\end{equation}
According to the formulas above, the difference between the input and label only lies at the markers' areas. Note that the difference brought by Poisson noise is mostly eliminated between the input and label volumes as well. The task-specific learning makes the network neglect unimportant structures and focus on marker restoration only, hence generating robust performance for complex clinical data.

\subsubsection{3D marker detection}
{With marker recovery, various existing marker detection algorithms can be applied.} In this work, the 3D {Hough} transform is applied for {spherical maker detection as an example.} Considering the large searching range in the reconstructed volume, a dynamic threshold is set according to an approximation of {the} pixel number of markers, which typically have {larger intensities compared to anatomical structures} in the recovered volume. Candidates whose {intensities are} above the threshold are chosen for sphere detection.




\subsection{Experimental Setup}

{The two proposed algorithms are trained from simulated data only, but are evaluated on simulated data and real data both for marker detection in navigation assisted spine surgery.
}

\subsubsection{Simulated data}
\begin{table}[htbp]
\centering
\captionv{8}{}{Parameters for the C-arm CBCT system and reconstruction.
\label{parameter}
}
    \begin{tabular}{l|l}
        \Xhline{2\arrayrulewidth}
        Parameter&Value\\
        \Xhline{2\arrayrulewidth}
         Scan angular range&200$^{\circ}$\\
        \hline
        Angular step&0.5$^{\circ}$\\
        \hline
        Source-to-detector distance& 1164 mm\\
        \hline
        Source-to-isocentor distance& 622 mm\\
        \hline
        Detector size& 976\,$\times$\,976\\
        \hline
        Extended virtual detector size& {2048}\,$\times$\,976\\
        \hline
        Detector pixel size & 0.305\,mm\\
        \hline
         FOV diameter & {16\,cm}\\
        \hline
       \hline 
        Reconstruction volume size&800\,$\times$\,800\,$\times$\,600\\
        \hline
        Reconstruction voxel size &0.313\,mm\\

        \Xhline{2\arrayrulewidth}
    \end{tabular}
\end{table}

In this work, 18 patients' CT data sets from the 2016 Low Dose CT Grand Challenge\cite{https://doi.org/10.1002/mp.12345} are used. Among them, 17 patients are used for training and one for test. {Spherical markers are inserted to each patient volume with their radii and intensities randomly chosen from 3\,-\,6\,mm and 1000\,-\,3000 HU, respectively. The number of markers is randomly chosen between 8 and 25 as one set. The markers have no overlap along the sagittal axis.} Because the markers are always located on the upper part of the image, to improve the accuracy of networks as well as reduce computation, the upper part of volumes is chosen for the network training.

Simulated projections are generated in a Siemens C-arm CBCT system with its system parameters reported in Table.~\ref{parameter}. Poisson noise is simulated 
by the rejection method \cite{10.2307/2346807} 
in the truncated projection data, {while untruncated projection data for reference reconstruction volumes contains no noise.} Water cylinder extrapolation \cite{https://doi.org/10.1118/1.1776673} is utilized to preprocess the acquired truncated projection data. The projection generation and image reconstruction are implemented in the framework of CONRAD \cite{https://doi.org/10.1118/1.4824926}.  

Considering different truncation and noise levels, the following three scenarios are investigated in this work:

(A) Regular: The regular scenario considers an FOV diameter of 16\,cm with the system parameters in Table.~\ref{parameter}, and a standard noise level assuming an exposure of 10$^6$ photons at each detector pixel before attenuation.
    
(B) Severer truncation: A smaller FOV diameter of 12\,cm with a standard noise level is considered.
    
(C) Heavier noise: A standard FOV diameter of 16\,cm with a heavier noise level assuming an exposure of 10$^4$ photons at each detector pixel before attenuation is considered.

\subsubsection{Parameters for {the} direct method}
\label{subsubsect:ParamtersDirectMethod}

{For the direct method, each patient's CT volume is inserted with one set of markers. For data augmentation, the marker insertion is repeated for 50 times. As a result, 850 CT volumes are used for training. For test, the marker insertion to the 18th patient's volume is repeated 10 times, where in total 166 markers are randomly generated in the 10 volumes.}
 
 For the in-plane coordinates, each volume with markers is compressed into 8 layers. Therefore, 6800 slices generated from 850 volumes are used for training the U-Net and among them 340 slices are used for validation. The Adam optimizer is used for optimization. The initial learning rate is 0.001 and the decay rate is 0.95. The loss function is mean absolute error. In total, the U-Net is trained for 100 epochs.

The 2D Hough transform is utilized for circle detection in the U-Net outputs. The search step is 0.5 pixel and the radius ranges are set between 5 and 10 pixels for simulated data. {For marker detection in real data, the radius is limited to a small range according to markers from different vendors. For example, 7\,-\,8 pixels are set to big markers and 4\,-\,5 pixels are set to small markers.}

For {the} depth information, the 850 volumes are also used to generate the training data. Since the positions of all the markers in the volumes are known, one of these markers is randomly chosen and its in-plane coordinates with a random bias of $\pm3$ pixels are set to be the axis for the cuboid, which has the size of {$400\,\times\,32\,\times\,32$}. Then a compression along {the} longitudinal axis is performed for {the corresponding} cuboid and one input image containing {a single marker} is generated and the corresponding label is divided by 400 for normalization. In total, 8000 images containing markers are generated randomly. For the {marker-free} situation, another 2000 images are generated from the 17 volumes without any modified markers and their labels are 0. The percentage of validation is $5\%$. The ResNet50 is trained for 200 {epochs}, with the loss function of mean squared error. The optimizer is Adam and the initial learning rate is 0.0001 with a decay rate of 0.999.
\subsubsection{Parameters for the recovery method}
\label{subsubsect:ParametersRecoveryMethod}

{To increase the sensitivity of neural networks in markers, hundreds of non-overlapping spherical markers instead of 8\,-\,25 markers are inserted above the spine of each patient. Otherwise, it is too slow to train the neural network if only 8-25 markers are in one volume. In total, 17 volumes are reconstructed for training. Due to the similarity between neighboring slices, one slice from every 5 slices (80 slices for each volume) is selected for training. For test, all the central 400 slices {of the 18th patient's reconstructed volume} are fed to the neural network.} For detection accuracy comparison with the direct method, we reuse the same 10 volumes of the 18th patient from Subsection.\,\ref{subsubsect:ParamtersDirectMethod}.

For network training, the Adam optimizer is utilized for optimization and its initial learning rate is 0.001 with an exponential decay rate of 0.95 for 500 decay steps. The values of $\beta_1$ and $\beta_2$ for Adam are 0.9 and 0.999, respectively. The mean absolute error is the loss function for the FBPConvNet. The FBPConvNet is trained on the mini batch with a batch size of 2, and it is trained for 100 epochs to get the final results. The generator in the Pix2pixGAN shares the common parameters above. Additionally, a weight $\alpha$ of 100 is used to combine the $\ell_1$ loss of the generator with the adversarial loss \cite{isola2017image}.

For the 3D Hough transformation, to reduce the searching range and accelerate the computation speed, a dynamic threshold for each real data volume in the upper part is calculated based the histogram. For example, for 7 big markers in real data with radius of 8 pixels, the threshold is the 15000th highest intensity in histogram. The step for searching is 1 pixel. The radii for big markers in real data volumes are 7 and 8 pixels and the radii for small markers are 4 and 5 pixels.
\subsubsection{Real data}
For real data experiments, the C-arm CBCT system parameters for data acquisition and image reconstruction are the same as those in Tab.\ref{parameter}. To evaluate the performance of the two algorithms, the following four real (cadaver) data sets are chosen as examples, considering data heterogeneity: 

(A) One thoracic volume with small markers: the markers have the same diameter of 3\,mm and the holder has 5\,cm distance to the patient's back.

(B) The second thoracic volume with big markers: the diameter of all the markers is 5\,mm and the holder is 5\,cm over the patient's back.

(C) One lumbar volume with big markers: all the markers have the diameter of 5\,mm and the holder is 5\,cm away from the patient's back. With more bones, this volume is more noisy and the intensity of markers is lower.

(D) The third thoracic volume with K-wires and big markers: the markers have the diameter of 5\,mm and the holder is placed directly at the back. The K-wires introduce metal artifacts.

\subsubsection{Evaluation {metrics}}
For marker recovery, the connected area of each marker with similar intensity can be {segmented} and {a} mean F1 score is calculated to see how well the recovered markers overlap with the ground truth {in simulated data}. What is more, the {intensity} difference between the prediction and {the corresponding reference is calculated}. For real data without reference, the orthogonal views and intensity plot of markers are {displayed} to provide the essential information for analysis.


{To provide some hints about the computational cost, the run-time evaluation of our implementation is displayed in Tab.\,\ref{summary}.} All codes are implemented in Python on a computer with Intel(R) Core(TM) i9-10900X CPU and a NVIDIA GeForce RTX 2080 SUPER graphics card.

\section{Results}
\subsection{Results of simulated data}

\subsubsection{Results of the direct method}
\begin{figure}[h]
   \centering
   \includegraphics[width=0.45\textwidth]{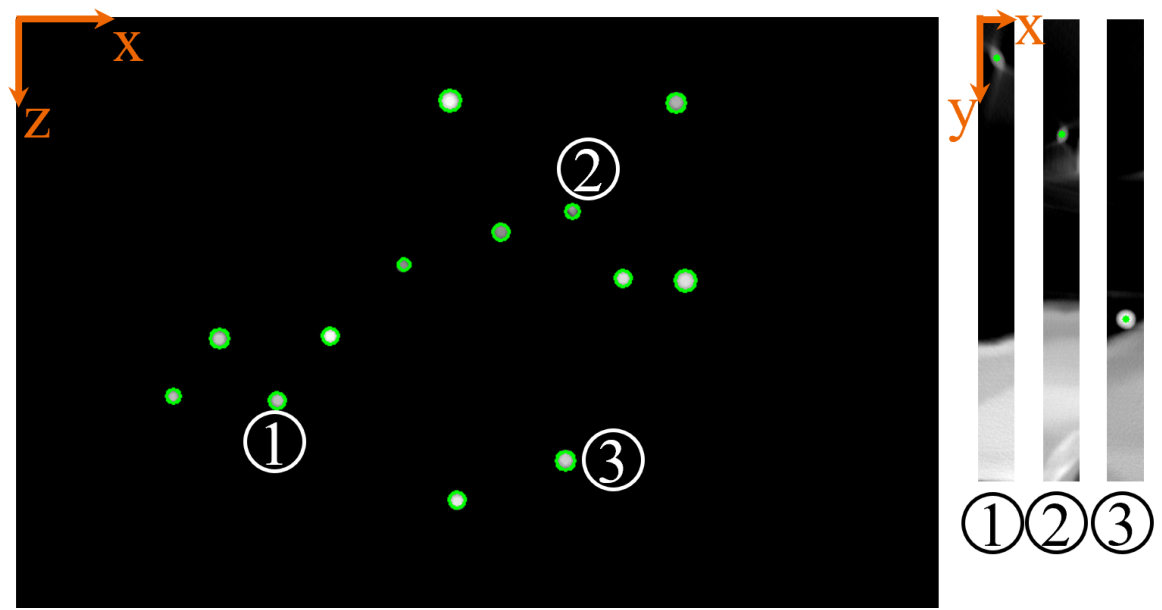}
   \captionv{8}{}
   {The marker detection results of the integrated output from {the} U-Net are displayed on the left side. The integrated rectangular images of three markers, labelled by No.\,1-3, for the ResNet50 are displayed on the right side and the detected y-coordinates for Maker No.\,1-3 are marked by the green dots.  
   \label{SimuDirectMethod}
    }
\end{figure}
The marker detection result for {an} integral image predicted from {the} U-Net is displayed in {Fig.\,\ref{SimuDirectMethod} as an example}. The detected markers are described by their circular borders in green in the x-z plane. The markers in Fig. \ref{SimuDirectMethod} {have} different {radii}, and all the markers are successfully detected.

Three detected circles, labelled by {No.\,1, 2, and 3}, are selected as examples {to further display} their corresponding rectangular images (on the right side of Fig.\,\ref{SimuDirectMethod}) generated by integral along {the} longitudinal axis. {The predicted y-coordinates by the ResNet50 are marked by the green dots. Fig.\,\ref{SimuDirectMethod} demonstrates that the ResNet50 is able to detect the y-coordinates very accurately. A quantitative evaluation of the overall detection accuracy on the simulated data is displayed in Subsection\,\ref{subsubsect:accuracyComparison}.}

\subsubsection{Results of the recovery method}

\begin{figure*}[hbt!]
   \centering
   \includegraphics[width=\textwidth]{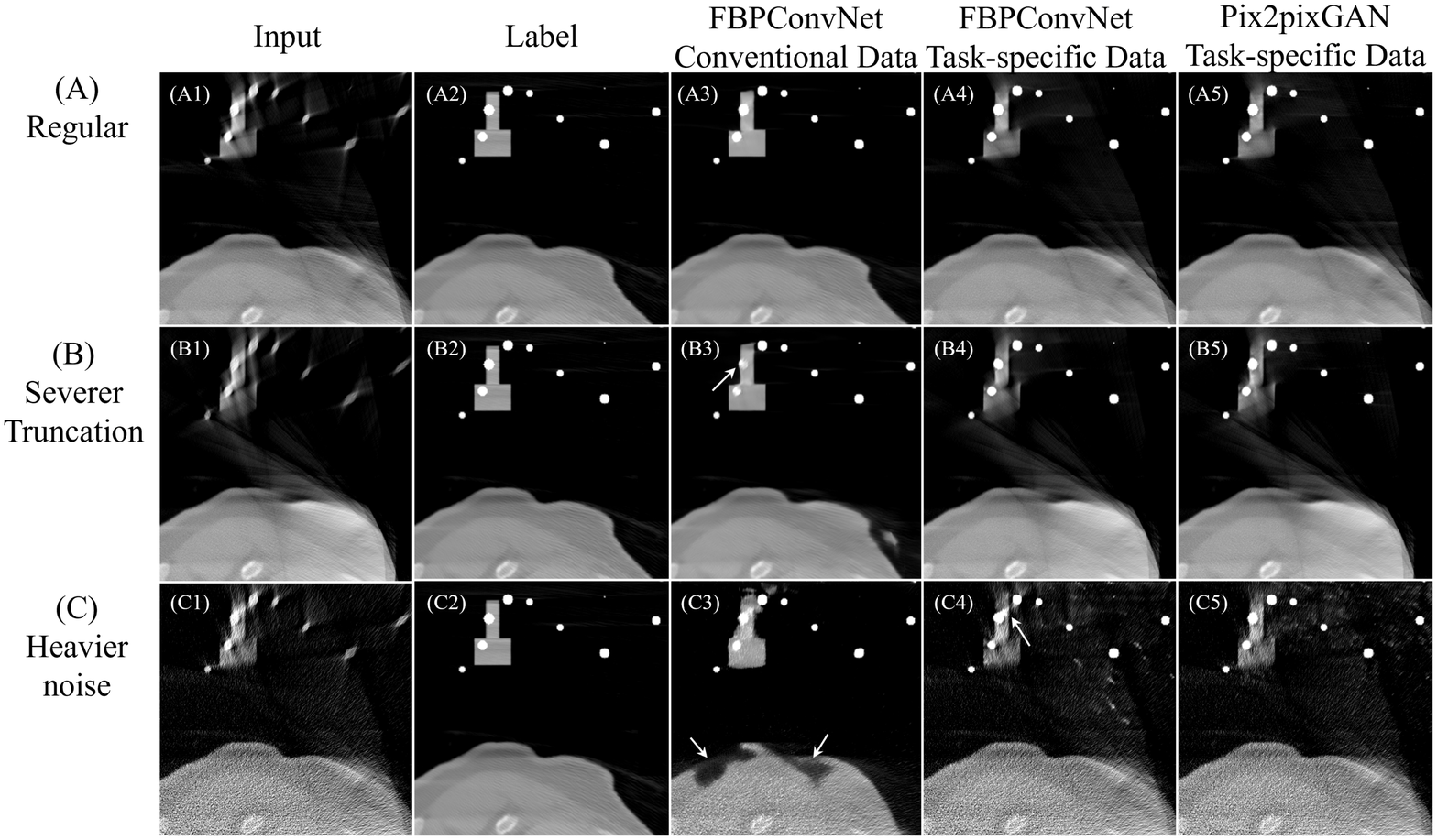}
   \captionv{16.5}{}
   {{Prediction} results by different neural networks on the three categories of simulated data: (A) Test data with {the} same noise level and truncation as training data; (B) Test data with {severer} truncation {than} training data; (C) Test data with {heavier} noise {than} training data, window: [-1000, 500]\,HU. 
   \label{SimulationData}
    }
\end{figure*}

\begin{table*}[hbt!]
\captionv{10}{}{\label{RecoveryCompa}Marker recovery comparison with reference}
\centering
\begin{tabular}{cp{0.8cm}p{0.8cm}p{0.05cm}p{0.8cm}p{0.8cm}p{0.05cm}p{0.8cm}p{0.8cm}p{0.05cm}p{0.8cm}p{0.8cm}}
\hline
\multirow{2}{*}{Test} &
\multicolumn{2}{c}{Input}& &
\multicolumn{2}{c}{FBPConvNet (I)}& &
\multicolumn{2}{c}{FBPConvNet (II)}& &
\multicolumn{2}{c}{{Pix2pixGAN} (II)}\\
\cline{2-3}
\cline{5-6}
\cline{8-9}
\cline{11-12}
  & $F_{1}$ & $d_{Intensity}$ && $F_{1}$ & $d_{Intensity}$ && $F_{1}$ & $d_{Intensity}$ && $F_{1}$ & $d_{Intensity}$\\
\hline
(A) Regular  & 78.4\% & 1006 && 97.8\% & 120 && 98.4\% & 76 && 98.6\% & 80\\
(B) Severer truncation  & 68.4\% & 1227 && 96.2\% & 237 && 97.3\% & 200 && 97.0\% & 216\\
(C) Heavier noise  & 75.5\% & 1025 && 93.7\% & 236 && 95.2\% & 245 && 96.0\% & 197\\
\hline
\multicolumn{12}{@{}l}{I--Conventional data\qquad II--Task-specific data\qquad $F_{1}$--F1 score\qquad $d_{Intensity}$--Mean Intensity difference (HU)}
\end{tabular}
\end{table*}
 
The prediction results {of the test patient's CT data in the three scenarios} are shown in Fig.~\ref{SimulationData}.
Three networks, {two FBPConvNets and one Pix2pixGAN}, are trained for comparison: the first FBPConvNet is trained on conventional data; the other FBPConvNet and {the Pix2pixGAN} are trained on task-specific data. For test data (A), which shares the same truncation and noise level as {the} training data, all three networks give comparable results {on marker recovery, as displayed in Figs.\,\ref{SimulationData}(A3)-(A5)}. However, for test data (B) or (C), {the} FBPConvNet trained on conventional data has degraded performance. It {predicts an} incomplete circle in Fig.~\ref{SimulationData}(B3) and brings distortion on anatomical structure in Fig.\,\ref{SimulationData}(C3). {In contrast, the} FBPConvNet trained on task-specific data is more robust, {as displayed in Fig.~\ref{SimulationData}(B4).} 
{Nevertheless, it fails to predict satisfying markers for the test data with heavier noise, as revealed by Fig.\,\ref{SimulationData}(C4). Instead, Pix2pixGAN trained from task-specific data has superior performance in all the three scenarios, as demonstrated by Figs.\,\ref{SimulationData}(A5)-(C5).}

{For image quality quantification}, the mean F1 score and brightness difference are calculated between the reference and prediction. In total 158 markers from the central 400 slices of the test patient's volume are taken into consideration. For each marker, the pixels with similar intensity are considered to {belong to} that marker's region. The comparison results are listed in Tab.\,\ref{RecoveryCompa}. The three test categories A, B and C correspond to the input data in Fig.\,\ref{SimulationData}. For the input data, test data B has the lowest F1 score and the biggest intensity difference compared with {the} reference because the markers have severer distortion. Test data {C} has little difference in F1 score and intensity compared to test data A. {For the output, the FBPConvNet trained from task-specific data has superior performance compared with the FBPConvNet trained from conventional data. Pix2pixGAN and FBPConvNet, both trained from task-specific data, have comparable performance in these three scenarios on simulated data.}
\subsubsection{{Accuracy comparison between two methods}}
\label{subsubsect:accuracyComparison}
\begin{figure}[hbt]
    \centering
    \begin{minipage}{0.94\linewidth}
    \subfigure[Detection with a false negative]{
    \includegraphics[width=\linewidth]{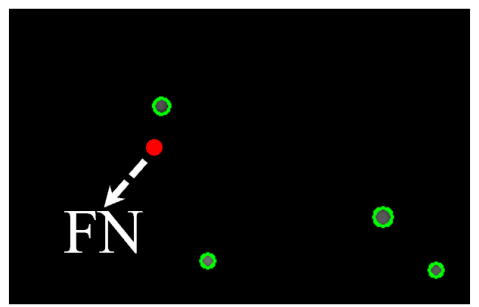}
    \label{subfig:FN_Example}
    }
    \end{minipage}
    \begin{minipage}{0.47\linewidth}
    \subfigure[Regular data]{
    \includegraphics[width=\linewidth]{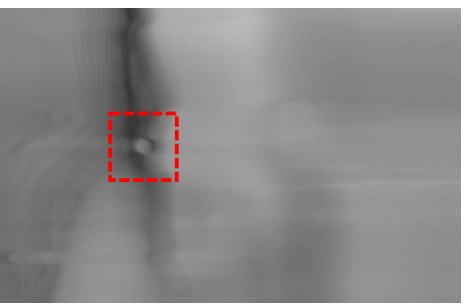}
    \label{subfig:FN_Input_R}
    }
    \end{minipage}
    \begin{minipage}{0.47\linewidth}
    \centering
    \subfigure[Severer truncation data]{
    \includegraphics[width=\linewidth]{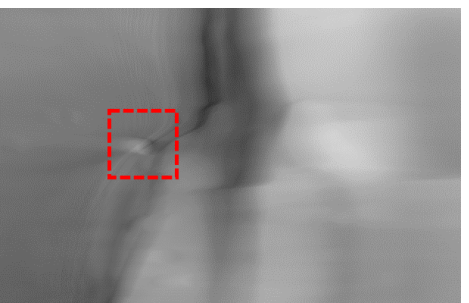}
    \label{subfig:FN_Input_ST}
    }
    \end{minipage}
 \captionv{8}{}{{A false negative (FN) example in the direct method: (a) is the upper left corner of the predicted integrate image of the U-Net, where the marker in red is not detected in severer truncation data set; (b) is the corresponding input data from the regular data set and the marker inside the red square is detected; (c) is the corresponding input data from the severer truncation data set and the marker inside the red square is not detected due to severer distortion.}}
  \label{Fig:FNDirect}
\end{figure}

\begin{figure}[hbt!]
 \centering
   \includegraphics[width=0.86\linewidth]{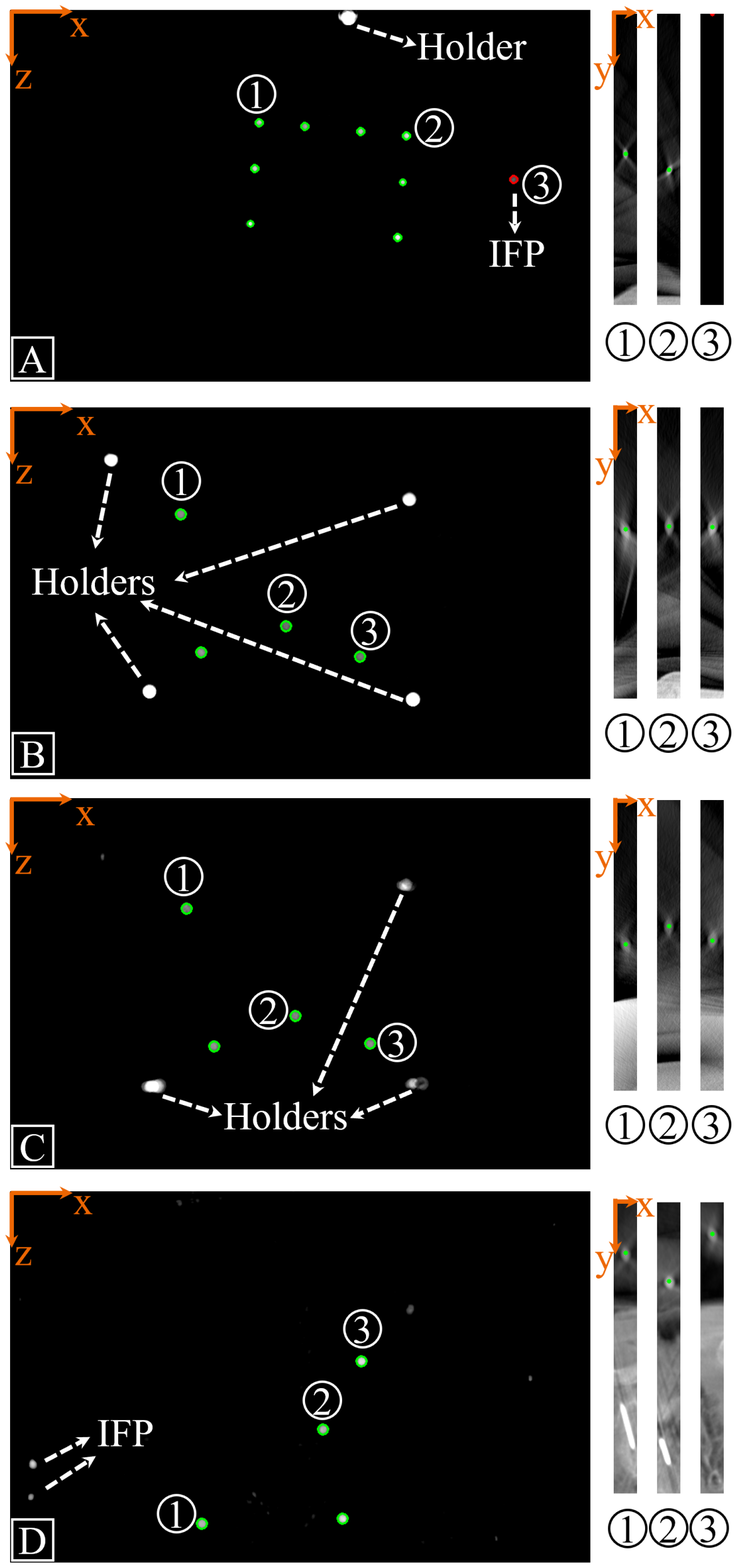}
   \captionv{8}{}
   {The marker detection results of integrated outputs from the U-Net for the four real data sets and several corresponding integrated rectangular images for the ResNet50. {The markers with green circles are final successfully detected markers. The bright regions in the x-z planes without green circles are intermediate false positive (IFP) detections (including the holders). The area marked by the red circle in (A) is an IFP detection by the conventional 2D Hough transform. The green dots on the right side mark the successfully detected y-coordinates.} 
   (A) Thoracic image with small markers; (B) Thoracic image with big markers; (C) Lumbar image with big markers;  (D) Thoracic image with K-wires and big markers. Not all the markers are displayed due to commercial reason. 
   \label{CliniDirectMethod}
    }
\end{figure}


\begin{table*}[hbt]
\captionv{8}{}{\label{Accuracy}Marker detection accuracy comparison on the three categories of simulated data.}
\centering
\begin{tabular}{p{0.3cm}cp{1cm}p{1cm}p{1cm}p{0.01cm}p{1cm}p{1cm}p{1cm}}
\hline
 \multirow{2}{*}{Test} & \multirow{2}{*}{$d$} &
\multicolumn{3}{c}{Direct method} &&
\multicolumn{3}{c}{Recovery method}\\
\cline{3-5}
\cline{7-9}
 && x & y & z && x & y & z\\
 \hline
 \multirow{3}{*}{$A$} &
 $=0$ & 76.5\% & 86.7\% & 71.1\% && 85.6\% & 89.8\% & 62.7\%\\
 & $\leq1$& 100\% & 99.4\% & 100\% && 100\% & 100\% & 100\%\\
 & $\leq2$& 100\% & 100\% & 100\% && 100\% & 100\% & 100\%\\
\hline
\multirow{3}{*}{$B$} &
 $=0$ & 51.8\% & 42.2\% & 65.1\% && 80.1\% & 87.3\% & 66.9\%\\
 & $\leq1$& 91.0\% & 92.8\% & 94.6\% && 100\% & 100\% & 100\%\\
 & $\leq2$& 94.0\% & 94.6\% & 94.6\% && 100\% & 100\% & 100\%\\
\hline
\multirow{3}{*}{$C$} &
 $=0$ & 73.4\% & 84.9\% & 69.9\% && 86.7\% & 90.4\% & 62.0\%\\
 & $\leq1$& 99.4\% & 99.4\% & 100\% && 100\% & 100\% & 100\%\\
 & $\leq2$& 100\% & 100\% & 100\% && 100\% & 100\% & 100\%\\
\hline
\multicolumn{9}{@{}l}{$d$--Pixel difference}
\end{tabular}
\end{table*}
The marker detection accuracy of the two methods are evaluated on simulated test data in the three scenarios (A)-(C).
The results of {the} recovery method are based on the predictions by Pix2pixGAN {with task-specific learning}. The marker detection accuracy comparison is listed in {Tab.\,\ref{Accuracy}}.

{Tab.\,\ref{Accuracy} shows that both the recovery method and the direct method have good performance on the standard data (A).} Only one marker is predicted to have {2-pixel deviation by the} direct method, {while all the markers detected by the recover method are within the 1-pixel precision}. With one voxel size of 0.313 mm$\,\times\,$0.313 mm$\,\times\,$0.313 mm, the detection accuracy is acceptable for clinical use. {For test data B (severer truncation), 
all the markers from the recovery method are within the 1-pixel precision along all the three dimensions,} but the direct method does not detect all the markers, with 5.4$\%$ markers undetectable {i.e., introducing false negative (FN) cases}. {In total, 9 markers out of 166 in ten volumes are missing, 8 of which are not detected by the U-Net and 1 is missed by the ResNet50. One example for a FN by the U-Net is displayed in Fig.\,\ref{Fig:FNDirect}.} {In test data C (heavier noise),} both methods are able to detect all the markers accurately within {the 2-pixel precision range}.

\subsection{Results of real data}

\subsubsection{{Results of the direct method}}

The results of {the} four real data sets are displayed in Fig.\,\ref{CliniDirectMethod}. The images {on the left side} show the predictions from the U-Net and the markers with green circles are the final successfully detected markers. The bright regions in the x-z planes without green circles are IFP detections (including the holders). The area marked by the red circle in Fig.\,\ref{CliniDirectMethod}(A) is an IFP detection by the conventional 2D Hough transform. The images on {the} right side are the integrals along {the} longitudinal axis for the detected circles with corresponding numerical labels, {and all the detected y-coordinates are marked by the green dots}. In Fig.\,\ref{CliniDirectMethod}(A)-(C), both the markers and holders with circular sections are segmented by {the} U-Net. But given the prior information of the marker radius, the IFP holders are removed by the 2D Hough transform in Fig.\,\ref{CliniDirectMethod}(A)-(C). However, in Fig.\,\ref{CliniDirectMethod}(A), one IFP circular area is detected by the 2D Hough Transform, {which is labelled} No.\,3. There is no marker existing in its corresponding integral image for the ResNet50, {and the ResNet50 predicts 0} for this marker-free situation. Therefore, this IFP detection by the 2D Hough transform is removed after the ResNet50. In Fig.\,\ref{CliniDirectMethod}(D), some IFP areas are segmented by the U-Net because of the existence of K-wires and their artifacts. But they are not detected by the 2D Hough transform because of their non-circular shapes.

\subsubsection{Results of the recovery method}

\begin{figure*}[htb!]
   \centering
   \includegraphics[height=0.78\textheight]{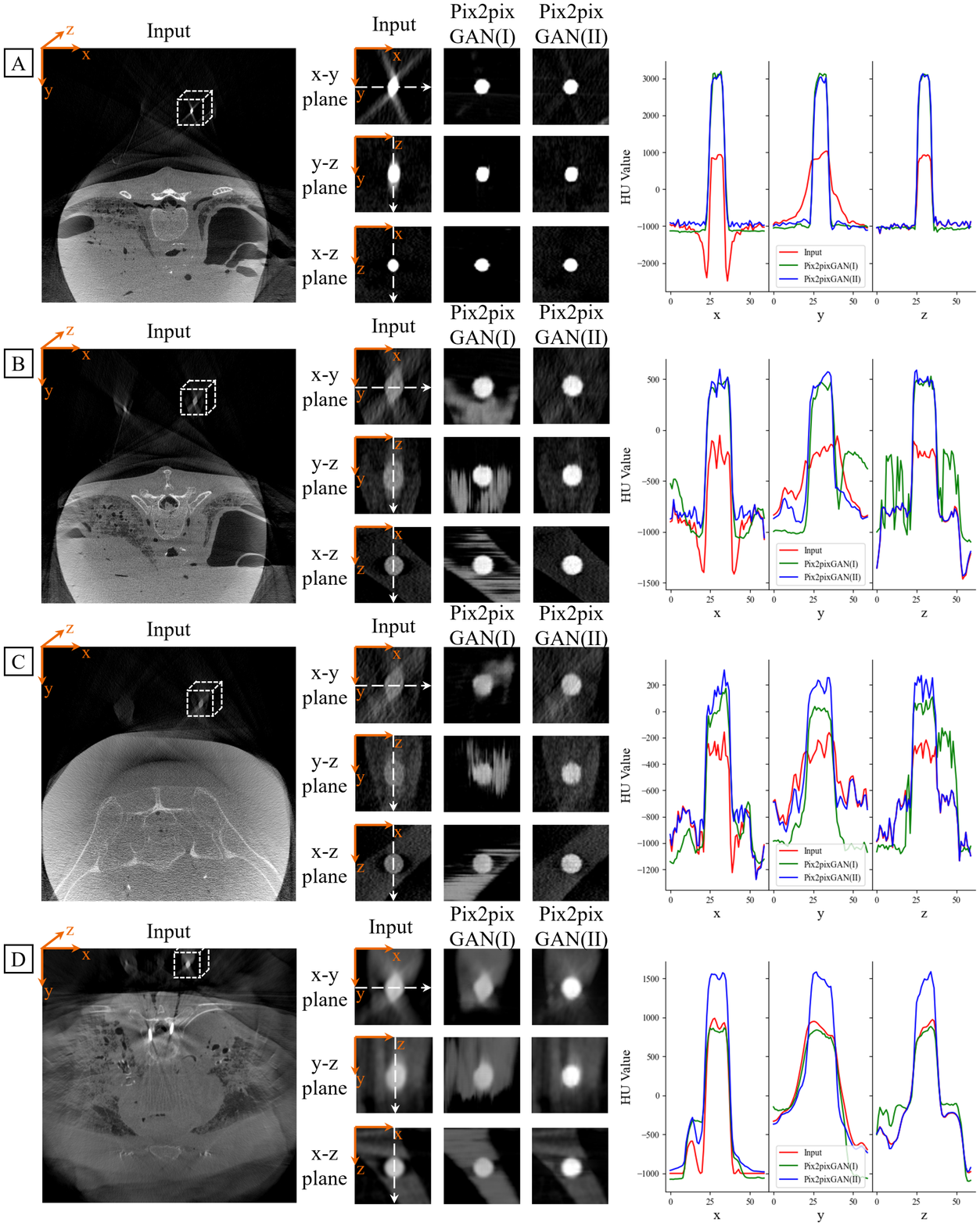}
   \captionv{16.5}{}
   {Prediction examples on the four real data sets, {window [-1000, 500]\,HU for (A)-(C) and window [-1000, 1500]\,HU for (D)}. Left column: an exemplary slice from the input volume for each data set is displayed. The referred directions ``x, y, z" are shown at the left top corner. A small cube is drawn in each input image, showing the range of the target marker to display. Middle column: the three orthogonal views for the selected marker in each data set are displayed to show shape restoration and {the recovery comparison between two Pix2pixGANs trained on conventional (I) and task-specific (II) data sets.} Right column: line intensity profiles (in HU) along each direction is plotted. The lines are marked in the corresponding middle column. A: Thoracic image with small markers; B: Thoracic image with big markers; C: Lumbar image with big markers; D: Thoracic image with K-wires and big markers. 
   \label{CombinedImages}
    }  
\end{figure*}
\begin{figure*}[t!]
  \centering
   \includegraphics[width=\textwidth]{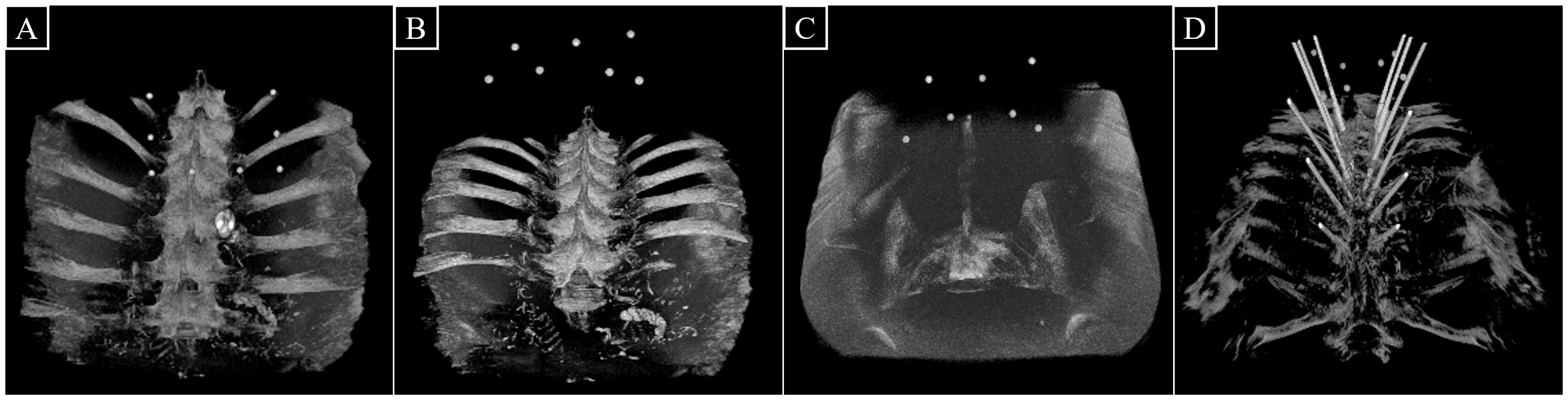}
   \captionv{16.5}{}
   {3D views for the predicted volumes by Pix2pixGAN {trained on task-specific data set} {on real data: (A) Thoracic image with small markers; (B) Thoracic image with big markers; (C) Lumbar image with big markers; (D) Thoracic image with K-wires and big markers. Not all markers are displayed due to commercial reason.}
   \label{3Dviews}
    } 
\end{figure*}

For the real data, the {results of} Pix2pixGAN trained on conventional data (Pix2pixGAN(I)) and task-specific data (Pix2pixGAN(II)) are {displayed in Fig.\,\ref{CombinedImages}}.
One exemplary input slice for each data set is displayed on the left side of Fig.\,\ref{CombinedImages}.  The referred directions ``x, y, z'' are shown at the left top corner of each input image. A small cube is drawn in each input image, showing the range of the target marker to display. Since there are no ground truth marker positions for real data, the three orthogonal views of each marker are displayed in the middle column of Fig.\,\ref{CombinedImages} for image quality assessment. 
For the input images of all these four data sets, shape distortion is clearly observed in the x-y and y-z planes, while there is little distortion in the x-z plane. For all the four data sets, Pix2pixGAN trained from both the conventional data and task-specific data can restore the marker shapes to a large degree, as indicated by the middle column figures. For (A), both Pix2pixGANs achieve comparable performance. For (B) and (C), a lot of bright artifacts are predicted by Pix2pixGAN trained from conventional data. For (C) and (D) in the x-y plane, Pix2pixGAN trained from task-specific data achieves better performance on shape restoration than the other.

To further quantify the marker intensities, line intensity profiles (in HU) along each direction is plotted on the right side of Fig.\,\ref{CombinedImages}. The red, green and blue curves stand for plots of input, prediction by Pix2pixGAN(I), and prediction by Pix2pixGAN (II), respectively. The three corresponding lines are marked as dotted white lines in the middle column.  {Regarding input images, along the x direction, because of dark streaks, there are drastic intensity drops near marker boundaries in the input images. {Nevertheless, the boundary position is preserved because of the sharp transition.} Along the y direction, due to the marker distortion, the background areas near the marker in the input images suffer from bright streak artifacts and thus have large intensities. As a consequence, the transition from a marker to the background is very smooth instead of being sharp. Along the z direction, the marker boundary in the input image is preserved and hence a sharp transition is observed. In all the three directions, intensity loss is observed. All the above phenomena in input images are clearly visible in Fig.\,\ref{CombinedImages}(A). Regarding the performance of the two Pix2pixGANs, for (A), both Pix2pixGANs are able to compensate the intensity loss in all the three directions as well as restore the marker boundary along the x and y directions. Consistent with the observations in the x-z plane on marker shapes, the line profiles along the z direction from the input and the two Pix2pixGANs overlap well at the marker boundaries. Both Pix2pixGANs only compensate the intensity for the marker areas. Among the four data sets, the line profiles for (B) and (C) have more fluctuations, indicating the severe noise level. As a result, the line profiles of Pix2pixGAN(I) for (B) and (C) have too large intensity values in the background areas. In (D), Pix2pixGAN(I) fails to compensate the marker intensity, as its output has approximately the same intensity as the input image. In contrast, Pix2pixGAN(II) is still able to compensate the marker intensity.}

It is worth noting that the existence of K-wires in (D) does not degrade the performance of Pix2pixGAN with task-specific learning. With high attenuation coefficient, they bring a lot of metal artifacts in the whole image, which can be a big obstacle for the networks. With task-specific learning, the Pix2pixGAN is still able to restore marker shapes as well as compensate the intensity loss.

The 3D views rendered by the ImageJ 3D Viewer for the four real data sets predicted by Pix2pixGAN trained from task-specific data are displayed in Fig.~\ref{3Dviews}. 
After setting a threshold 0 HU, artifacts around markers and most soft tissues are gone. These markers can then serve as a bridge for the registration between volume coordinates and real world coordinates.

\subsubsection{Marker registration comparison between two methods}
 \begin{table}[htb!]
 \captionv{8}{}{\label{Registrationerror}Marker position difference after registration for two methods}
 \centering
 \begin{tabular}{p{0.3cm}p{0.3cm}p{1.2cm}p{1.2cm}p{1.2cm}p{1.2cm}}
 \hline
 \multirow{2}{*}{} & 
 \multirow{2}{*}{$N$} &
 \multicolumn{2}{c}{Recovery method} &
 \multicolumn{2}{c}{Direct method}\\
 \cline{3-6}
  && $\bar e$ (mm) & $\sigma$ (mm) & $\bar e$ (mm) & $\sigma$ (mm)\\
 \hline
  A & 12 & 0.101 & 0.033 & 0.101 & 0.042\\
  B & 7 & 0.117 & 0.057 & 0.165 & 0.080\\
  C & 7 & 0.188 & 0.122 & 0.120 & 0.056\\
  D & 7 & 0.103 & 0.030 & 0.196 & 0.065\\
 \hline
 \multicolumn{6}{@{}l}{$N$--Number of markers\qquad $\bar e$--Mean error}\\
 \multicolumn{6}{@{}l}{$\sigma$--Standard deviation}
 \end{tabular}
\end{table}
The real data has no reference on image volume, thus a registration is conducted for these markers between the accurate positions provided by vendors. In the four categories of real data, all the markers are successfully detected by both methods. The mean {FRE} and deviation for markers in each volume after marker alignment and registration are listed in Tab.\,\ref{Registrationerror}. 
It shows that both methods achieve the mean difference for markers in all four volumes smaller than 0.2 mm and the deviation within markers in each volume {about 0.1 mm}.

\section{Discussion}
\begin{table*}[hbt!]
\captionv{16.5}{}{\label{summary}Summary for two methods}
\centering
\begin{tabular}{ccccc}
\hline
& Component & Function & Time &{Overall robustness}\\
\hline
\multirow{3}{*}{Direct method} & U-Net & Marker extraction & 15 s & \multirow{3}{*}{FN risk and FP tolerance} \\
& 2D Hough tansform & 2D position detection & 1 s &\\
& ResNet50 & Depth detection & 6 s  & \\
\hline
\multirow{2}{*}{Recovery method} & pix2pixGAN & Marker recovery & 42 s  &\multirow{2}{*}{{FN and FP tolerance}}\\
& 3D Hough tansform & 3D position detection & {15 s}  &\\
\hline
\end{tabular}
\end{table*}
{The direct method contains three steps for marker position detection directly in CBCT volumes reconstructed from severely truncated data. In the output of the U-Net, IFP circular areas might be segmented since deep learning solely is not robust. Since a conventional circle detection method with prior constraints, i.e. the 2D Hough transform with the prior information of the marker radius in this work, is integrated in the direct method, IFP circular areas can be eliminated to a large degree, as demonstrated by Fig.\,\ref{CliniDirectMethod}(B)-(D). Even though some IFP areas remain after the 2D Hough transform, they can be removed by the ResNet50 at the last step, as demonstrated by Fig.\,\ref{CliniDirectMethod}(A).}

The ResNet50 is capable of dealing with this {marker-free} situation, because the images without markers are also generated for the training process. {Because of the above mentioned multi-step mechanism, our direct method has tolerance to FP cases in general.}
{However, in severer truncation situation, the direct method has a risk of FN detection cases, as indicated by Tab.\,\ref{Accuracy}(B) where} 5.4\% markers are not successfully detected. Among the 9 missing markers, 8 are missed by the U-Net and 1 is missed by the ResNet50. According to their coordinates, those markers are mainly distributed near {the} boundaries. Therefore, those markers have {much severer} distortion and looks like a long oval in the integral image. Because the training data does not have such examples, {the U-Net and the ResNet50 fail to deal with such out-of-the-distribution samples}. But for clinical images, markers are mainly distributed right above patients' spines {which do not have so severe distortion as that in Fig.\,\ref{subfig:FN_Input_ST}.} Therefore, all the markers in real data volumes are successfully detected by the direct method.

{The instability of deep learning methods is a major concern for their clinical applications. The amount and distribution of training data as well as noise are common factors to influence the robustness and generalizability of deep learning methods \cite{huang2018some,antun2020instabilities}. Because of this, incorrect structures are predicted by deep learning models, for example, the IFP circles in the intermediate result in Fig.\,\ref{CliniDirectMethod}(A)-(D), and the incorrect anatomical structures in Fig.\,\ref{SimulationData}(C3). With our proposed task-specific learning strategy, the neural network focuses on SOI only, and hence can extract essential features related to the task only instead of being disturbed by other features from unimportant structures. For example, in the marker recovery application,}  compared with the conventional data preparation strategy, the difference between the input and label data is much smaller and it only lies in the markers' region. With unaltered structures in the remaining image, the network focuses on the recovery of markers but not on the truncation correction for anatomy or noise reduction. {Because of this, although some contexts in the test input are not contained in the training data, e.g. the existence of K-wires (Fig.\,\ref{CombinedImages}(D)) or heavier noise (Fig.\,\ref{SimulationData}(C)), the neural network is able to ignore them and still generate stable outputs.}

A summary for a comparison of {the} two methods is listed in Tab.\,\ref{summary}. {Despite the risk of FN cases in the severer truncation situation, the direct method is able to detect the positions of markers for real data, as demonstrated by Fig.\,\ref{CliniDirectMethod}. In addition, it is computationally efficient since it processes small 3D subvolumes and 2D patches only. Considering robustness and detection accuracy (Tab.\,\ref{Accuracy}) as well as} the maturity of conventional algorithms for marker detection, the recovery method is superior. {With a 3D volume processed by our task-specific learning, the custom conventional marker detection algorithms from various vendors can be simply plugged in.}

Data truncation is a common problem for CBCT systems with flat-panel detectors, which limits their applications. The state-of-the-art deep learning methods have limited performance for FOV extension from severely truncated data. The results of the fiducial marker recovery in spine surgery 
demonstrate that our proposed task-specific learning can empower CBCT systems with many more potential applications, which are currently impossible because of the severe truncation. It indicates a promising prospect of our proposed task-specific learning in applications with severely truncated data in the near future.

\section{Conclusion}
This study {focuses} on the automatic detection of markers from severely truncated data for universal navigation assisted {MISS}, so that the imaging system works well with navigation systems from different vendors. {The multi-step direct detection method and} the recovery method with task-specific learning are explored. The direct method has better computation efficiency in marker detection, succeeding in all real data cases. {With the multi-step mechanism, it is able to remove 
IFPs. However, it may have a risk of FNs for marker detection with severer truncation than usual.}
With task-specific learning and pix2pixGAN, {the recovery} method is robust in different real data volumes and the combined 3D {Hough} transform has high marker detection accuracy with maximal 1 pixel difference in all coordinates. The task-specific learning has the potential to reconstruct SOI accurately from severely truncated data, which might empower CBCT systems with new applications in the near future.

\textbf{Disclaimer:} The concepts and information presented in this paper are based on research and are not commercially available


\section*{References}
\addcontentsline{toc}{section}{\numberline{}References}
\vspace*{-10mm}





{\footnotesize

}     





\end{document}